\documentclass[prc,twocolumn,APS,showpacs,preprintnumbers,amsmath,amssymb,superscriptaddress,floatfix]{revtex4}

\usepackage{graphicx}
\usepackage{dcolumn}
\usepackage{bm}

\newcommand{\fmn}[2]{\mbox{${\textstyle \frac{#1}{#2}}$}}

\begin{document}

\title{Differential cross section and analyzing power of the
$\roarrow{p}p{\to}pp{\pi}^0$ reaction at a beam energy of
390\,MeV}

\author{Y.\,Maeda}
\email{ymaeda@rcnp.osaka-u.ac.jp}%
\affiliation{Research Center for Nuclear Physics, Osaka University,
Ibaraki, Osaka 567-0047, Japan}%
\affiliation{Institut f{\"u}r Kernphysik, Forschungszentrum
J{\"u}lich, 52425 J{\"u}lich, Germany}
\author{M.\,Segawa}
\altaffiliation[Present address: ]{Nuclear Science Research
Institute, Tokai Research and Development Center,
Tokai-mura, Naka-gun, Ibaraki 319-1195, Japan  }
\affiliation{Research Center for Nuclear Physics, Osaka
University, Ibaraki, Osaka 567-0047, Japan}
\author{T.\,Ishida}
\altaffiliation[Present address: ]{Laboratory of Nuclear Science,
Tohoku University, Sendai, Miyagi 982-0826, Japan.}
\affiliation{Department of Physics, Kyushu University, Fukuoka
812-8581, Japan}
\author{A.\,Kacharava}
\affiliation{Institut f{\"u}r Kernphysik, Forschungszentrum
J{\"u}lich, 52425 J{\"u}lich, Germany}
\affiliation{High Energy Physics Institute, Tbilisi State
University, 0186 Tbilisi, Georgia}%
\author{M.\,Nomachi}
\affiliation{Department of Physics, Osaka University, Toyonaka,
Osaka 560-0043, Japan}
\author{Y.\,Shimbara}
\altaffiliation[Present address: ]{National Superconducting
Cyclotron Laboratory, Michigan State University, East Lansing,
Michigan 48824-1321, USA
}%
\affiliation{Department of Physics, Osaka University, Toyonaka,
Osaka 560-0043, Japan}
\author{Y.\,Sugaya}
\affiliation{Department of Physics, Osaka University, Toyonaka,
Osaka 560-0043, Japan}
\author{K.\,Tamura}
\affiliation{Physics Division, Fukui Medical University, Fukui
910-1193, Japan}
\author{T.\,Yagita}
\affiliation{Department of Physics, Kyushu University, Fukuoka
812-8581, Japan}
\author{K.\,Yasuda}
\affiliation{The Wakasa Wan Energy Research Center, Fukui 914-0192, Japan}
\author{H.P.\,Yoshida}
\altaffiliation[Present address: ]{
Cyclotron and Radioisotope Center, Tohoku University, Sendai, Miyagi 980-8578, Japan
}%
\affiliation{Research Center for Nuclear Physics, Osaka
University, Ibaraki, Osaka 567-0047, Japan}
\author{C.\,Wilkin}
\affiliation{Physics and Astronomy Department, UCL, Gower Street,
London WC1E 6BT, United Kingdom}

\date{\today}

\begin{abstract}
The differential cross section and analyzing power $A_y$ of the
$\roarrow{p}p{\to}pp{\pi}^0$ reaction have been measured at RCNP
in coplanar geometry at a beam energy of 390\,MeV and the
dependence on both the pion emission angle and the relative
momentum of the final protons have been extracted.
The angular variation of $A_y$ for the large values of
the relative momentum studied here shows that this is primarily an
effect of the interference of pion $s$- and $p$-waves and
this interference can also explain the momentum dependence.
Within the framework of a very simple model, these results
would suggest that the pion-production operator has a
significant long-range component.
\end{abstract}


\maketitle

%
%
\section{\label{sec:INTRO}Introduction}
\setcounter{equation}{0}
Pion production is the first and probably the simplest of the
inelastic processes in nucleon-nucleon collisions and its
understanding may provide us with valuable information about low
and medium energy strong interaction physics. Motivated by this,
over the last decade there has been a series of detailed studies
of the $pp{\to}pp{\pi}^0$ reaction from near the production
threshold up to a proton beam energy of 425\,MeV. These have been
carried out at several laboratories, \textit{viz.}\
IUCF~\cite{pi0_iucf,pi0_iucf_new},
TSL~\cite{pi0_tsl,pi0_tsl_ang,Pia}, and COSY-TOF~\cite{TOF1,TOF2}.

Initial theoretical calculations~\cite{Miller} for
$\pi^0$-production estimated a total cross section that was five
times smaller than experiment~\cite{pi0_iucf}. This indicated
clearly that some essential mechanism was missing from the theory
of $s$-wave pion production leading to the $S$-state of the final
protons. Lee and Riska~\cite{Lee} suggested that this might be
connected with short-range effects between nucleons, and
quantitative support for this was found in a model with the
exchange of a heavy meson coupled to the antinucleon-nucleon
pair~\cite{Horowitz}. On the other hand it has been claimed that
$s$-wave pion rescattering is not small~\cite{Oset,Juelich} and
that one can reproduce the cross section data without invoking
short-range effects~\cite{Oset}. Studies based on Chiral
Perturbation Theory ($\chi$PT)~\cite{Park,Cohen,Kolck,Sato} show
that the contribution of pion rescattering is indeed sizable, but
that the sign of this term is opposite to that of
Ref.~\cite{Oset,Juelich}, and this leads to an even more severe
discrepancy between theory and experiment. The convergence of the
chiral expansion can be seriously questioned for $s$-wave pion
production because of the large momentum transfer between
nucleons, whereas the expansion seems to show convergence in the
case of $p$-wave pion~\cite{HanhartCHIPT}. An ordering scheme has
since been discussed \cite{HanhartKaiser}. Thus the origin of
$s$-wave pion production is still not clear even as to whether the
production mechanism is or is not dominated by short-range
effects.

Differential cross sections and polarization observables have been
measured at several bombarding
energies~\cite{pi0_iucf_new,pi0_tsl_ang,TOF2,Pia} and theoretical
estimates of these observable have been made by groups at
J{\"u}lich~\cite{Juelich2} and Osaka~\cite{tamura}. Both
approaches include higher partial-waves and provide good fits to
the total cross section close to threshold. However the
differential and polarization observables are not well reproduced
by either group. In order to try to identify the origin of the
problem more clearly it may be useful to attempt a partial-wave
decomposition. In Ref.~\cite{pi0_iucf_new}, the angular
dependences of the different polarization observables were
developed using a general formalism consisting of a complete set
of functions, with coefficients extracted by fitting data, and
this has been extended in later work~\cite{Deepak}. However, due
to the limited accuracy of polarization data existing at the time,
the analysis was done by assuming that only a small number of
partial-wave contributed and that the momentum dependence of these
amplitude stemmed purely from the centrifugal barrier. This latter
assumption is very doubtful since, for the $P$-state of final
protons, it does not lead to a good description of the
differential cross section~\cite{pi0_tsl_ang}. More precise data
on the angular and momentum dependence of the polarization
observables, as well as of the differential cross section, are
highly desirable. The accumulation of such data may allow one to
perform a fuller partial-wave analysis of the $pp{\to}pp{\pi}^0$
reaction, which would provide greater insight for the theoretical
models.

We report here a measurement of the $\roarrow{p}p{\to}pp{\pi}^0$
reaction at a beam energy of 390\,MeV, where the beam polarization
is perpendicular to the plane containing the detectors. The
differential cross section and analyzing power, $A_y$, are
obtained as functions of the pion emission angle, $\theta_q$, in
the center-of-mass system and the relative momentum, $p$, of the
final protons. The use of a high intensity polarized beam in
conjunction with a liquid hydrogen target makes it possible to
extract the angular dependence of $A_y$ for different ranges in
$p$. Such data, which have not been available previously, might
offer serious constraints on theoretical models.

If only a few partial-waves are important, the angular dependence
of the $A_y$ $\times$ spin-averaged cross section is composed of
terms proportional to $\sin\theta_q$ and $\sin2\theta_q$. The
strength of  the $\sin\theta_q$ term is governed by the
interference of amplitudes corresponding to the production of $Ps$
and $Pp$ final states. Here, using the standard notation, the
final states are labeled by $Ll$, where $L$ and $l$ are the
angular momenta of the proton pair and the pion, respectively. On
the other hand, the $\sin2\theta_q$ strength is determined by $S$-
and higher states.

The variation with the momentum $p$, obtained after integration
over angles, is sensitive to the $Ps$ and $Pp$ production
amplitudes. This is studied in a model where it is assumed that
the momentum dependence of the proton-proton wave functions in the
final state is the main influence for the $P$-state amplitudes.
Furthermore, in a very simplistic approach, the wave functions are
evaluated at a \textit{fixed} proton-proton separation distance.

The experimental facility, and in particular the detectors,
polarized beam, and target, are described in Sec.~\ref{sec:EXP}.
Section~\ref{sec:ANALYSIS} is devoted to the steps needed to
identify and measure the $pp{\to}pp{\pi}^0$ reaction. The extraction
of the observables from data taken under our specific kinematic
conditions, where the detection system is essentially coplanar, is
the subject of Sec.~\ref{Defkinematics}, with the experimental
results being shown in Sec.~\ref{Results}. The general features of
the angular and momentum dependence of the cross section and
analyzing powers are given in Sec.~\ref{discussion}. It is shown
there that, when comparing our results with published data which
have large acceptance, it is crucial to take account of our
particular coplanar geometry. Section~\ref{phenomenal} presents
the simple phenomenological description of pion production. Our
conclusions are given in Sec.~\ref{summary}.
%
%
\section{Experiment}\label{sec:EXP}
\begin{figure*}[htb]
\includegraphics[clip,width=0.8\textwidth]{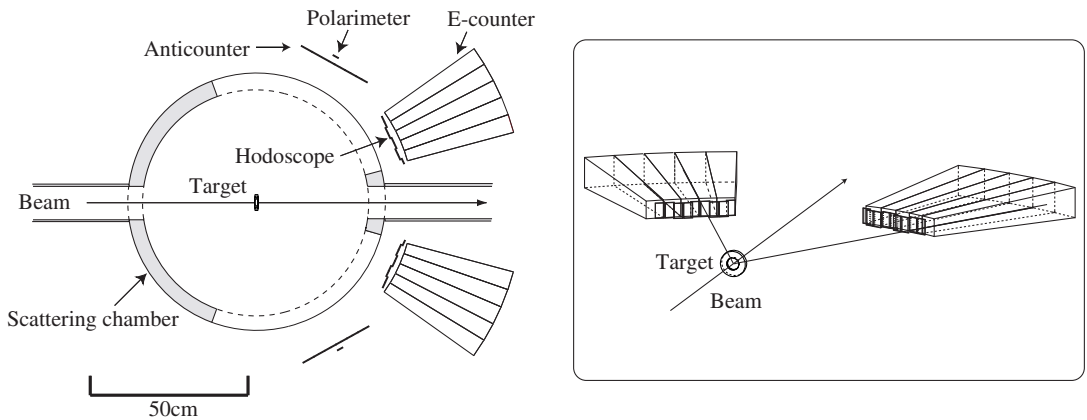}
\caption{\label{fig:setup} Top view of the layout of the
experiment. The shaded region shows the walls of the target
chamber and the dashed lines its windows. In the right-hand panel
a three-dimensional sketch shows the positions of the hodoscope
and $E$-counters. }
\end{figure*}

The experiment was carried out using a 390 MeV polarized proton
beam extracted from the cyclotron complex at the Research Center
for Nuclear Physics (RCNP), Osaka. The polarized beam was produced
by an atomic-beam-type polarized ion source with an Electron
Cyclotron Resonance (ECR) ionizer~\cite{ECR}, where the
polarization state (`up' or `down') of the primary beam was
reversed with a frequency of 1\,Hz. The protons were first
accelerated in the injector AVF cyclotron before being further
accelerated up to 390 MeV in the main cyclotron ring. After
extraction, the vertically polarized beam was transported to the
scattering chamber in the experimental hall.
Figure~\ref{fig:setup} shows the top view of the scattering
chamber as well as of our detection system. The scattering chamber
containing a liquid hydrogen target consisted of an evacuated
vertical cylinder. The window made of a aramid foil had a
horizontal opening angle from 15$^\circ$ to 110$^\circ$ on either
side of the the beam direction. After the target the primary beam
was transported to the beam dump where a Faraday cup monitored the
beam intensity for the two polarization states. The beam current
was limited to around 1\,nA in order to minimize the dead-time of
the data acquisition system. Under these conditions the system,
the details of which are described in Ref.~\cite{DAQ}, had an
efficiency of between 75\% and 87\%.
%
%
\subsection{\label{sec:DETECTOR}Detector}
The detector system was symmetric with respect to the plane
containing the beam axis and the direction of the beam
polarization. The two outgoing protons from the $pp{\to}pp{\pi}^0$
reaction were detected simultaneously in an array of plastic
scintillators. The measurement of two pairs of polar and azimuthal
angles and two kinetic energies is sufficient to identify the
$pp{\to}pp{\pi}^0$ reaction and determine the five independent
kinematic variables. The system covered laboratory polar angles
$15^{\circ}<\theta<35^\circ$. The maximum angle of protons from
the $pp{\to}pp{\pi}^0$ reaction at 390\,MeV ($32.1^\circ$) is well
inside the angular acceptance. The minimum angle was limited for
both protons such that only proton pairs with relative momenta
from 150\,MeV/$c$ up to the kinematic limit of 220\,MeV/$c$ could
be registered with this system. The detector was therefore well
suited for the investigation of the high relative momentum region,
where the pion-production amplitudes leading to $P$-state proton
pairs should have their maximum strength. However, it is important
to note that the c.m.\ polar angle of the relative momentum vector
for final protons was confined to $90^{\circ}\pm30^{\circ}$ with
respect to the beam axis.

The energy of a scattered proton was deduced from the amplitude
signal in one of the $E$-counters. Such a counter consists of a
set of five plastic scintillators in which each element has a
trapezoidal shape, with front and back faces of area
$35\times35\,\textrm{mm}^2$ and $60\times60\,\textrm{mm}^2$
respectively. The 350\,mm length is sufficient to stop protons
with energies up to 250\,MeV so that all protons from the
$pp{\to}pp{\pi}^0$ reaction at 390\,MeV were stopped. The front
face of each counter was positioned at 500\,mm from the target.
The scintillator hodoscopes in front of the counters were used to
determine the directions of outgoing particles; a single 3\,mm
thick element of area $17.5\times30\,\textrm{cm}^2$ covered
$\pm$17\,mrad horizontally and $\pm$30\,mrad vertically. 
Two elements were placed in front of each of four $E$-counters with a
1\,mm overlap, though only one element was used for the counter 
at the largest angle. 
There was thus a total of nine elements on each side of the
beam. The angle of the detected particle was defined by the center
of each element.

The anticoincidence counter ( marked `Anticounter' in
Fig.~\ref{fig:setup}) was used to eliminate accidental events
coming from elastic proton-proton scattering. This covered the
angular range $45^{\circ}-72^{\circ}$, where recoil protons from
the elastic scattering hit the counter in combination with the
scintillator hodoscopes. The information from the counter was used
in the offline analysis and about 90\% of the accidentals could be
suppressed in this way.

One hodoscope element and a scintillation counter placed at
60$^{\circ}\pm$1$^{\circ}$ (denoted as the `Polarimeter' in
Fig.~\ref{fig:setup}) were used to monitor the beam polarization.
The fast and recoil protons from elastic proton-proton scattering
were detected in coincidence. According to the SAID
database~\cite{said}, the $pp{\to}pp$ analyzing power at this
energy and angle is $A_y=-0.364\pm0.007$, where the error bar has
been obtained by looking at typical data in this region. The
average values of the beam polarizations deduced on this basis
were $P_{\uparrow}=0.66\pm0.01$ and $P_{\downarrow}=0.70\pm0.01$,
where the arrows indicate the spin states of proton beam and the
errors are statistical. The polarizations were quite stable during
the experiment and all values lay within about $\pm 0.03$ of the
averages.
%
%
\subsection{\label{sec:TARGET}Target}
The liquid hydrogen (LH$_2$) target was 8.5\,mm thick, with
windows made of 25$\,\mu$m thick aramid foil~\cite{target}. Its
temperature was controlled in the range of 14\,K to 20\,K and kept
stable at $17.6\pm 0.4\,$K during experimental runs. The
\textit{empty target} runs, which were carried out with hydrogen
gas at temperatures between 22\,K and 27\,K, were used to estimate
the background from other construction materials as well as from
the residual gas frozen on the windows.
%
%
\subsection{\label{sec:DATATAKING}Data taking}
A signal was registered when a charged particle hit both of the
scintillator hodoscopes and the corresponding $E$-counter. For the
measurement of the $pp{\to}pp{\pi}^0$ reaction, the trigger
conditions were set so that both right- and left-side counters
were required to give signals in coincidence. In addition,
measurements without the coincidence requirement were also
performed in order to detect single protons from elastic
proton-proton scattering. Such measurements were also performed
using an unpolarized proton beam in order to check the
determination of the luminosity as well as of the beam
polarization. These are the subject of the following section.
%
%
\section{\label{sec:ANALYSIS}Data analysis}
%
%
\subsection{Particle identification}
\begin{figure}[hbt]
\includegraphics[clip,width=0.45\textwidth]{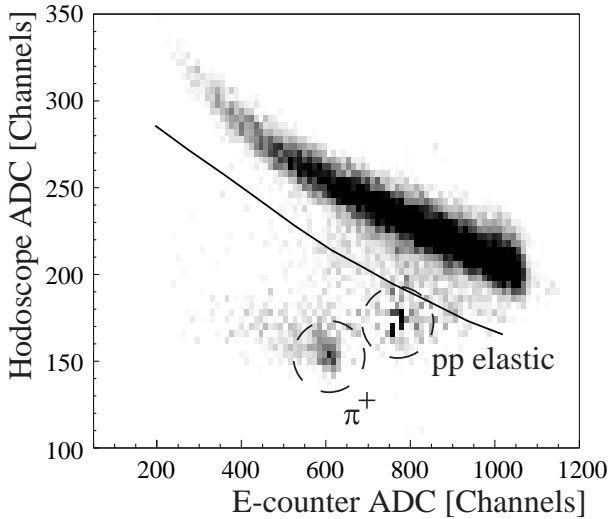}
\caption{\label{fig:deltae} Identification of protons from the
$pp{\to}pp{\pi}^0$ reaction. The signal from the hodoscope is
plotted versus that from the $E$-counter. The intense band is due
to protons that have stopped in the scintillator. Groups of events
corresponding to elastically scattered protons and pions from the
$pp{\to}pn\pi^+$ reaction are indicated and separated by the cut
shown as the solid line.}
\end{figure}

Particle identification was achieved by the $\Delta{E}$-$E$
method. The amplitude signal, related to the energy loss in the
scintillator hodoscope, is plotted in Fig.~\ref{fig:deltae}
against the amplitude signal from the $E$-counter. The heavily
populated band arises from protons, associated mostly with $\pi^0$
production, that stopped inside the counter. The $pp{\to}pp\pi^0$
events of interest are well separated from the two lower islands.
These are produced by elastically scattered protons, which have
energies above 265\,MeV and thus do not stop inside the counter,
and positive pions generated through the $pp{\to}pn{\pi^+}$
reaction.

%
%
\subsection{\label{sec:EnergyCalib}Energy calibration of the $E$-counter}
At a beam energy of 390\,MeV, the energies of the detected protons
from the $pp{\to}pp{\pi}^0$ reaction varied between 40\,MeV and
226\,MeV. The energy dependence of the amplitude signal from the
$E$-counter could be calibrated above 100\,MeV using elastic
proton-proton scattering events. The hodoscope and $E$-counter
were set to cover the angular range $40^{\circ}-60^{\circ}$ and,
for an angle fixed by the hodoscope, the monoenergetic recoil
proton was measured by an $E$-counter element. The energy
resolution was found to be better than 2\% (FWHM) for a proton
energy of 200\,MeV.

\begin{figure}
\includegraphics[clip,width=0.35\textwidth]{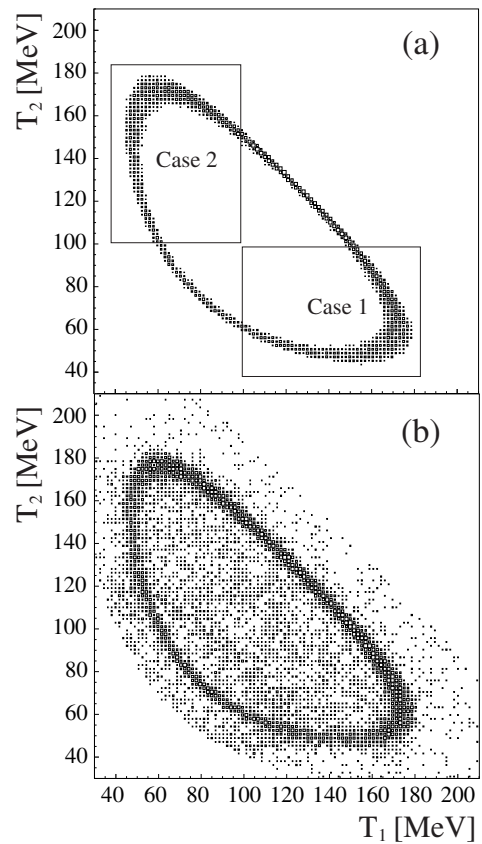}
\caption{\label{fig:tl1tr1} Two-dimensional plot of the proton energy
$T_2$ deposited in the right-hand $E$-counter versus the same
quantity $T_1$ for the left-hand counter. The hodoscope elements,
 located closest to the beam line on both sides, were
selected. Panels (a) and (b) show the simulated and experimental
data, respectively.}
\end{figure}

Below 100\,MeV, information from the $pp{\to}pp{\pi}^0$ reaction
itself was used. Figures~\ref{fig:tl1tr1}a and \ref{fig:tl1tr1}b
show, respectively, the simulated and measured energy correlation
between protons that hit a pair of hodoscope elements to the right
and left of the beam. The locus of $\pi^0$ event can be clearly
seen and this was used to extract a data sample where the energy
in the left (right) counter was higher than 100\,MeV, while that
in the right (left) was lower, as indicated by Case 1(2) in
Fig.~\ref{fig:tl1tr1}a. In addition a pedestal value from the ADC
module was also used at an energy of 17\,MeV, corresponding to the
minimum energy of protons detected in the $E$-counter. In this
way, the amplitude signals for all combinations of right and
left-side counters were calibrated using third-order polynomials
with an accuracy of better than 4\%, as judged from the position
and width of the resulting $\pi^0$ missing-mass peak.
%
%
\subsection{\label{sec:IDPI0}Identification of the $\mathbf{pp{\to}pp\pi^0}$ reaction}
\begin{figure}
\includegraphics[clip,width=0.4\textwidth]{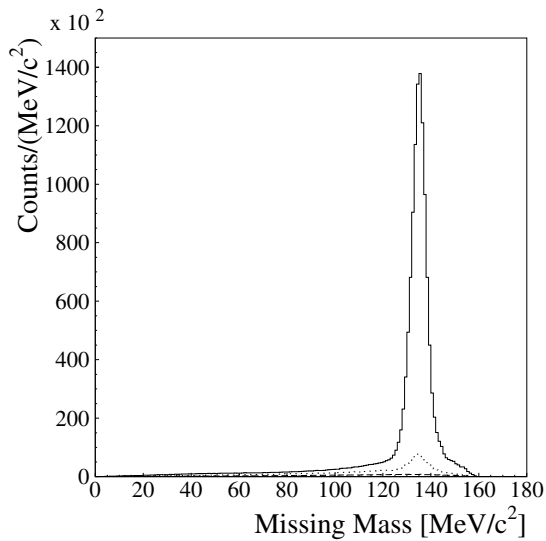}
\caption{\label{fig:totmass} The $pp{\to}ppX^0$ missing-mass
spectrum obtained for all geometric configurations using the
LH$_2$ (solid histogram) and the gaseous hydrogen target of the
`empty run' (dotted line). The estimate of random coincidence
events is shown by the dashed line. The peak corresponding to
$\pi^0$ production has a mass of 135.0\,MeV/$c^2$ and a width of
7.4\,MeV/$c^2$ (FWHM).}
\end{figure}

Having measured the energies and angles of the two protons, good
events were selected on the basis of the missing-mass spectrum. A
clear $\pi^0$ peak, with a width of 7.4\,MeV/$c^2$ (FWHM), is seen
in Fig.~\ref{fig:totmass}, which shows the totality of events
obtained with the LH$_2$ target. The background from the target
foil was subtracted by utilizing the \textit{empty target} runs
normalized to the integrated beam intensity. The contribution from
the random coincidences was estimated by considering outgoing
protons from different beam bunches. The correspondence of the
peak position with the mass of the $\pi^0$ is consistent with the
energy calibration of the counters.

For all $pp{\to}pp{\pi}^0$ candidates, the pion emission angle
$\theta_q$ and the relative momentum $p$ and its angle
$\theta_p$ in the overall center-of-mass system, were
reconstructed on an event-by-event basis. The data were then
grouped in nine intervals in $\theta_q$ and three in $p$, with a
finer binning in $p$ being used to study the momentum dependence
of the analyzing power.

\begin{figure}
\includegraphics[clip,width=0.45\textwidth]{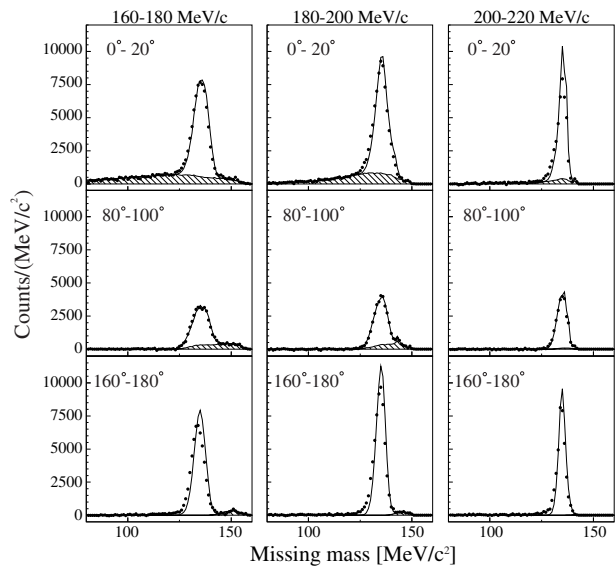}
\caption{\label{fig:submass} The $pp{\to}ppX^0$ missing-mass
spectra for different selected kinematic regions, with the
relative momentum $p$ being indicated at the top of the panels and
the angle of the pion $\theta_q$ being shown within each. The
experimental data are presented after subtracting the backgrounds
from the random coincidences and target window. The hatched
regions show normalized distributions of `non-full' events, where
not all of the energy is deposited in the scintillator. The solid
lines show the sums of these plus the good events. The shapes of
both distributions have been determined through simulations with
GEANT3~\cite{GEANT}}
\end{figure}

Missing-mass spectra were constructed for each combination of bins
and spin states and used to extract the yield of ${\pi}^0$ events.
The backgrounds from random coincidences and target foils were
subtracted before showing the data in Fig.~\ref{fig:submass} for
typical conditions. There is some residual background for
$\theta_q<20^{\circ}$, whose shape varies with the kinematic
conditions. This arises from `non-full' events where, in contrast
to `full' events, recoil protons from ${\pi}^0$ production do not
deposit their full energy in the $E$-counter due to a nuclear
reaction in the scintillator material.

The shapes of the distributions for non-full and full events were
simulated in a Monte Carlo program based on GEANT3~\cite{GEANT},
which took into account the energy resolution of the scintillator
material, as well as hadronic and electromagnetic interactions.
The simulated events were passed through the same analysis chain
as the measured data and fitted to the results shown in
Fig.~\ref{fig:submass}. The normalization of the non-full events
was determined from the tails of the distributions. In order to
check the systematic uncertainty in the treatment of the non-full
background, the $\pi^0$-yield was also determined by selecting all
events around the pion peak without any subtraction and the
corresponding uncertainty is discussed in Sec.~\ref{Results}.
%
%
\subsection{\label{sec:FULLEFF}Detection efficiency of $E$-counters for `full' events}
The detection efficiency for full events, where protons deposit
their full energy inside an $E$-counter, was estimated using the
data obtained for the energy calibration of the $E$-counters
discussed in Sec.~\ref{sec:EnergyCalib}. The efficiency varies
between 0.85 and 0.65 over the measured energy range of 100\,MeV
to 200\,MeV. The efficiency obtained by the Monte-Carlo simulation
was checked by comparing it with that obtained from the
measurement. Both results are in agreement to within 15\% and this
is included in the overall uncertainty quoted for the cross
section.
%
%
\subsection{\label{sec:LUMINOSITY}Luminosity}
\begin{figure}
\includegraphics[clip,width=0.48\textwidth]{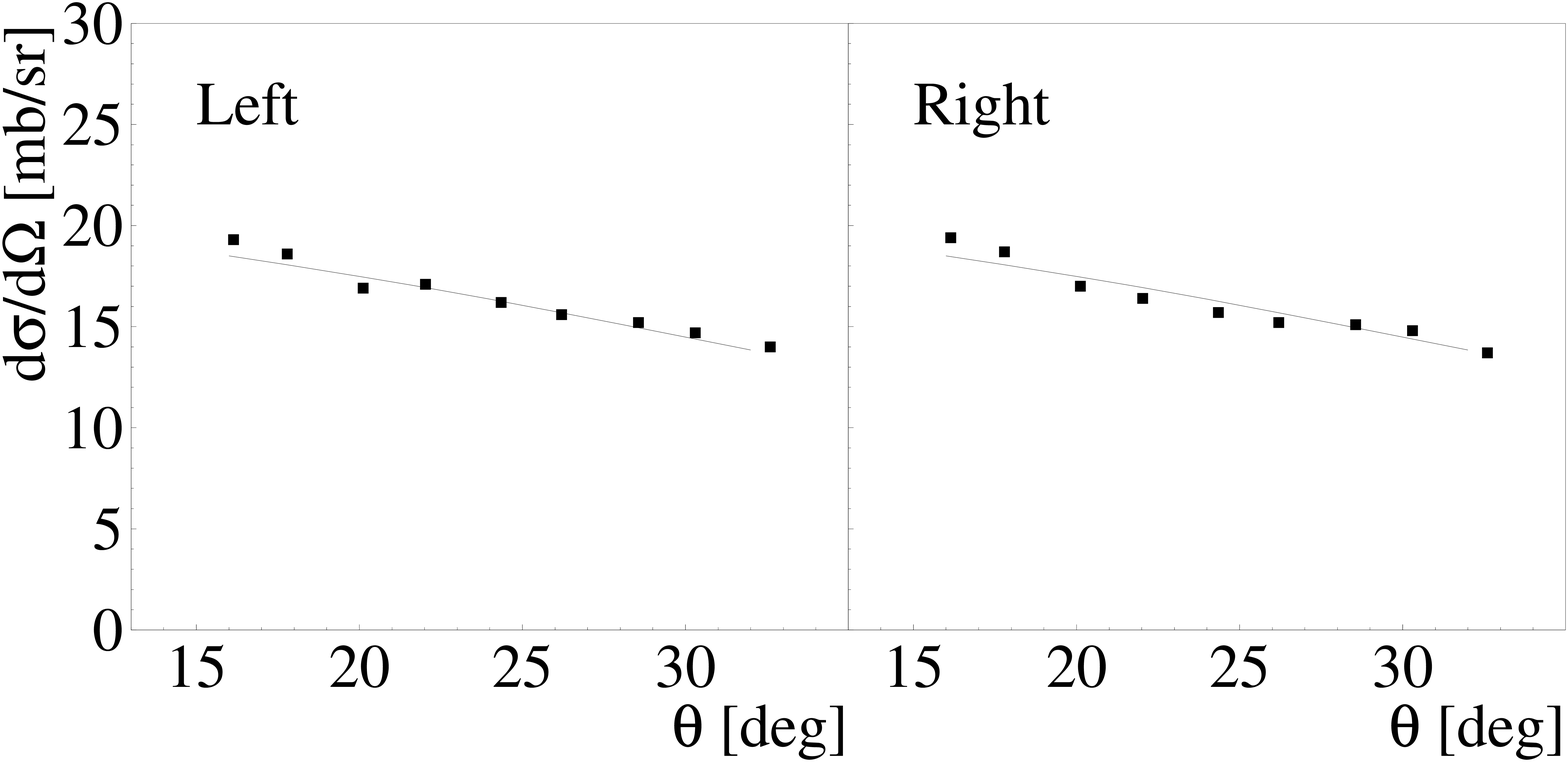}
\caption{\label{fig:ppelastic} The laboratory differential cross
sections measured for elastic proton-proton scattering using hodoscopes
placed to the left and right of the beam. The lines are
predictions taken from the SAID database~\cite{said}. }
\end{figure}

The absolute value of the luminosity was determined from
measurements of the beam intensity and the density of the LH$_2$
target and this gave an integrated luminosity of
$16.1\,\textrm{pb}^{-1}$. The value was verified by measuring
elastic proton-proton scattering by identifying single protons
through the $\Delta{E}$-$E$ method. The unpolarized $pp{\to}pp$
differential cross sections obtained in this way are shown
separately in Fig.~\ref{fig:ppelastic} for the same angular
interval to the left and right of the beam direction. The results
agree on average to within $\pm2\%$ with the predictions of the
SAID program~\cite{said}. However, there are fluctuations of up to
$\pm 5\%$ around the predicted curves and these deviations were
included in the determination of the acceptance as corrections
associated with the geometrical uncertainties of individual
hodoscope elements.
%
%
\subsection{\label{sec:BPol}Beam polarization}
\begin{figure}[t]
\includegraphics[clip,width=0.48\textwidth]{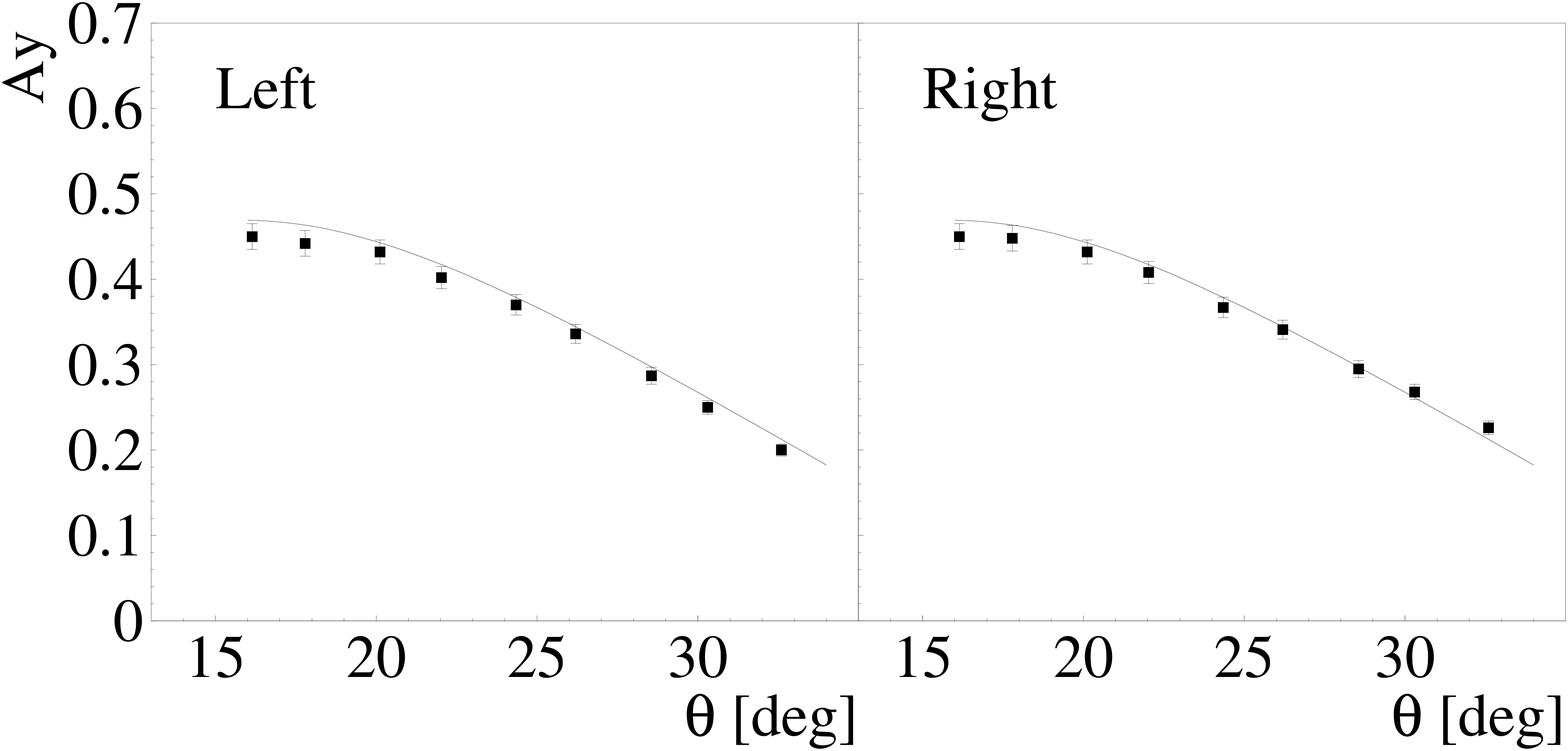}
\caption{\label{fig:ppelasticay} The analyzing power for elastic
proton-proton scattering measured by detecting single protons with
hodoscopes placed to the left and right of the beam, using beam
polarizations determined by the coincidence method. The lines are
estimates taken from the SAID database~\cite{said}. }
\end{figure}

The beam polarizations were also checked by measuring elastic
proton-proton scattering where the selection criteria for a single
proton is the same as that discussed previously. The analyzing
powers obtained in the offline analysis on the basis of the
already determined beam polarizations $P_{\uparrow,\downarrow}$
are shown in Fig.~\ref{fig:ppelasticay} for the same laboratory
angular range to the left and right of the beam direction. The
results are consistent with those given by the SAID
program~\cite{said} and the systematic uncertainty in the
determination of the beam polarizations, including that coming
from the SAID database, was found to be below 6\%.

%
%
\section{Observables}\label{Defkinematics}
\subsection{Definition}
%
%
Five independent variables $\xi$ are required to describe the
three-body final state and these we take from the relative
momentum $\mathbf{p}$ of the two final protons and that of the
pion $\mathbf{q}$ in the overall c.m.\ system. Since the
magnitudes of these momenta are linked by energy conservation, the
resulting set of variables consists of the magnitude of
$\mathbf{p}$ plus two polar angles $\theta$ with respect to the
beam axis and two azimuthal angles $\phi$, \textit{i.e.}\
$\xi\equiv$\{$\theta_q,\varphi_q,\theta_p,\varphi_p,p$\}. The
corresponding differential cross section will be denoted
\begin{equation}
\sigma(\xi, P_y ) \equiv \frac{d\sigma}{d\Omega_q d\Omega_p
dp}\:\cdot \label{eq:defcr}
\end{equation}

In terms of the Cartesian observables, the dependence of the cross
section on the vertical polarization $P_y$ is given by
\begin{equation}
\sigma(\xi, P_y ) = {\sigma}_0(\xi) ~[ ~1 + P_y A_y(\xi) ~]
~,\label{eq:defpolcr}
\end{equation}
where ${\sigma}_0$ is the spin-averaged cross section.

Our goal is to obtain the spin-averaged cross section and
analyzing power at the polar angle of $\theta_p = 90^{\circ}$ in
coplanar geometry, \emph{i.e.} $\varphi_p= \varphi_q=0^{\circ}$.
These observables are obtained from the experimental data through
\begin{eqnarray}
\label{eq:sigmaex} {\sigma}_0 &=&
\frac{\left(P_{\downarrow}L_{\downarrow} N_{\uparrow} +
P_{\uparrow} L_{\uparrow} N_{\downarrow}\right)} {\epsilon
L_{\uparrow}L_{\downarrow}(P_{\downarrow} +
P_{\uparrow})}~,\\
A_y &=& \frac{ (L_{\downarrow} N_{\uparrow} - L_{\uparrow}
N_{\downarrow}) / \langle \cos \varphi_q \rangle}
{\left(P_{\downarrow} L_{\downarrow} N_{\uparrow} + P_{\uparrow}
L_{\uparrow} N_{\downarrow}\right) }~, \label{eq:ayex}
\end{eqnarray}
where $N_{\uparrow,\downarrow}$ is the spin-dependent yield,
$P_{\uparrow,\downarrow}$ the beam polarization, and
$L_{\uparrow,\downarrow}$ the luminosity. The detection efficiency
$\epsilon$ includes that of the data acquisition system and the
acceptance of the detector system, as determined by the
Monte-Carlo simulation using the phase-space model. The average
value of the cosine of the pion angle, $\langle \cos \varphi_q
\rangle$, is also determined through the Monte-Carlo simulation,
as discussed in Sec.~\ref{Results}.
%
%
\subsection{Angular dependence}
For the later discussion it is convenient here to describe the
angular dependence expected for the unpolarized cross section and
analyzing power of the $pp{\to}pp{\pi}^0$ reaction. These, as well as
other observables, have been discussed in terms of partial-wave
amplitudes in Ref.~\cite{pi0_iucf_new}, where total orbital
angular momentum excitations up to $L_{pp}+\ell_{\pi}=2$ were
considered. Taking $\varphi_p=0^{\circ}$ and $\theta_p =
90^{\circ}$, the relevant formulae reduce to
\begin{eqnarray}
\sigma_0(\theta_q,\varphi_q,p) &=& E +
F_1+H^{00}_0-\left(H^{00}_2+F_2+K\right)+\nonumber \\
&& \hspace{-3cm} \left(I + H^{00}_1 -H^{00}_3\right)
(3\cos^{2}{\theta_q}-1)
 +H^{00}_5 \sin^2\theta_q \cos2\varphi_q\:,\nonumber \\
\label{eq:sigma_rcnp}\\
\sigma_0 A_y(\theta_q,\varphi_q,p)&=& \left\{\left(G^{y0}_1 -
G^{y0}_2+G^{y0}_4\right)\sin\theta_q\right. \nonumber \\
&&\hspace{-2cm} \left. +\left(I^{y0}+H^{y0}_1- H^{y0}_2 +
H^{y0}_5\right) \sin2\theta_q
\right\} \cos\varphi_q\:,\nonumber\\
\label{eq:ay_rcnp}
\end{eqnarray}
where each of the coefficients is a function of the relative
momentum $p$. The coefficients $E$, $F$, and $H$ represent the
absolute-squares of amplitudes leading to $Ss$, $Ps$, and $Pp$
states, respectively, whereas those with $G$, $I$, and $K$ reflect
the interferences of type $PsPp$, $SsSd$, and $SsDs$
respectively~\cite{pi0_iucf_new}.
%
%
\section{Results}
\label{Results}
\begin{table*}
\caption{\label{tab:ayang_table1} 
The spin-averaged $pp{\to}pp{\pi}^0$
 cross section $\sigma_0$ and analyzing power $A_y$
measured in different ranges of relative momentum $p$ at the polar
angle of $\theta_p = 90^{\circ}$ with its angular bin of $\pm
30^{\circ}$. The first and second error bars in $\sigma_0$ and
$A_y$ show, respectively, the statistical and systematic
uncertainties, whereas that for the pion emission angle $\theta_q$
is the standard deviation that includes the effects of the
binning. The overall systematic uncertainty for $A_y$ is estimated
to be less than 6\% and that for $\sigma_0$ to be less than 16\%
and these are not included in the systematic uncertainties for an
individual angular bin. The observables have been obtained at the
average values of $\langle\cos2\varphi_q\rangle$, whereas the
value of $\langle\cos\varphi_q\rangle$ is involved in the
evaluation of $A_y$. In order to get estimates of the values at
$\cos2\varphi_q=1$, the cross section should be divided by the
correction factor $d_t$ and the analyzing power multiplied by the
same factor, as discussed in the text.\\ }
\begin{ruledtabular}
\begin{tabular}{ccccccc}
${\theta}_q$&$p$& $\sigma_0$& $A_y$ & $\langle\cos\varphi_q\rangle$ & $\langle\cos2\varphi_q\rangle$ & $d_t$
\\
(deg) & (MeV/$c$) & \big(nb\,sr$^{-2}$\,(MeV/$c$)$^{-1}$\big)&&&&\\
 \hline
 $\phantom{1}14\pm 7$&&$ 7.64\pm 0.03\pm 0.16 $&$-0.085\pm 0.008\pm 0.005 $&$ 0.653 $&$ \phantom{-}0.206 $&$1.01$\\
 $\phantom{1}31\pm 7$&&$ 7.26\pm 0.03\pm 0.19 $&$-0.191\pm 0.006\pm 0.018 $&$ 0.950 $&$ \phantom{-}0.810 $&$1.01$\\
 $\phantom{1}50\pm 7$&&$ 5.84\pm 0.02\pm 0.23 $&$-0.259\pm0.007\pm0.005$&$ 0.977 $&$ \phantom{-}0.910 $&$1.02$\\
 $\phantom{1}70\pm 7$&&$ 4.89\pm 0.03\pm 0.38 $&$-0.345\pm 0.009\pm 0.047 $&$ 0.984 $&$ \phantom{-}0.938 $&$1.02 $\\
 $\phantom{1}90\pm 7$&$160-180$&$ 4.92\pm 0.03\pm 0.23 $&$-0.362\pm 0.018\pm 0.066 $&$ 0.985 $&$ \phantom{-}0.943 $&$1.02$\\
$110\pm 7$&&$ 5.24\pm 0.03\pm 0.13 $&$-0.307\pm 0.011\pm 0.018 $&$ 0.982 $&$ \phantom{-}0.933 $&$1.02$\\
$130\pm 7$&&$ 5.95\pm 0.03\pm 0.09 $&$-0.270\pm 0.009\pm 0.022 $&$ 0.972 $&$ \phantom{-}0.896 $&$1.02$\\
$148\pm 7$&&$ 7.32\pm 0.03\pm 0.31 $&$-0.208\pm 0.007\pm 0.009 $&$ 0.941 $&$ \phantom{-}0.779 $&$1.02$\\
$166\pm 7$&&$ 7.57\pm 0.03\pm 0.10 $&$-0.084\pm 0.009\pm 0.005 $&$ 0.590 $&$ \phantom{-}0.101 $&$1.01$\\
\hline
$ \phantom{1}16\pm 8$&&$ 5.56\pm 0.02\pm 0.18 $&$-0.076\pm 0.011\pm 0.024$&$0.536 $&$\phantom{-}0.072 $&$1.01$\\
$ \phantom{1}32\pm 8$&&$ 5.26\pm 0.02\pm 0.18 $&$-0.167\pm 0.006\pm 0.004$&$0.905 $&$\phantom{-}0.679 $&$1.02$\\
$ \phantom{1}51\pm 8$&&$ 4.49\pm 0.02\pm 0.03 $&$-0.259\pm 0.007\pm 0.005$&$0.959 $&$\phantom{-}0.843 $&$1.02$\\
$ \phantom{1}71\pm 8$&&$ 3.74\pm 0.02\pm 0.16 $&$-0.279\pm 0.010\pm 0.023$&$0.972 $&$\phantom{-}0.890 $&$1.03$\\
$ \phantom{1}90\pm 8$&$180-200$&$ 3.38\pm 0.02\pm 0.14 $&$-0.339\pm 0.008\pm 0.024$&$0.974 $&$\phantom{-}0.902 $&$1.03$\\
$109\pm 8$&&$ 3.69\pm 0.02\pm 0.12 $&$-0.310\pm 0.008\pm 0.029$&$0.971 $&$\phantom{-}0.886 $&$1.03$\\
$129\pm 8$&&$ 4.08\pm 0.02\pm 0.19 $&$-0.215\pm 0.007\pm 0.006$&$0.956 $&$\phantom{-}0.833 $&$1.03$\\
$148\pm 9$&&$ 5.09\pm 0.02\pm 0.12 $&$-0.150\pm 0.006\pm 0.007$&$0.904 $&$\phantom{-}0.661 $&$1.02$\\
$163\pm 8$&&$ 5.76\pm 0.02\pm 0.02 $&$-0.096\pm 0.009\pm 0.006$&$0.561 $&$\phantom{-}0.043 $&$1.01$\\
\hline
$ \phantom{1}22\pm 14$&&$ 2.34\pm 0.02\pm 0.06 $&$-0.114\pm 0.031\pm 0.042 $&$0.366 $&$-0.129$&$1.02$\\
$ \phantom{1}37\pm 12$&&$ 2.04\pm 0.02\pm 0.05 $&$-0.143\pm 0.017\pm 0.010 $&$0.725 $&$\phantom{-}0.303$&$1.03$\\
$ \phantom{1}54\pm 12$&&$ 1.77\pm 0.02\pm 0.02 $&$-0.194\pm 0.015\pm 0.011 $&$0.833 $&$\phantom{-}0.533$&$1.04$\\
$ \phantom{1}72\pm 11$&&$ 1.57\pm 0.01\pm 0.09 $&$-0.253\pm 0.016\pm 0.025 $&$0.881 $&$\phantom{-}0.646$&$1.05$\\
$ \phantom{1}91\pm 11$&$200-220$&$ 1.48\pm 0.01\pm 0.08 $&$-0.255\pm 0.016\pm 0.038 $&$0.896 $&$\phantom{-}0.677$&$1.05$\\
$109\pm 11$&&$ 1.54\pm 0.01\pm 0.12 $&$-0.233\pm 0.015\pm 0.028 $&$0.894 $&$\phantom{-}0.664$&$1.05$\\
$127\pm 13$&&$ 1.67\pm 0.01\pm 0.13 $&$-0.168\pm 0.015\pm 0.006 $&$0.848 $&$\phantom{-}0.546$&$1.04$\\
$143\pm 13$&&$ 1.89\pm 0.02\pm 0.10 $&$-0.164\pm 0.016\pm 0.011 $&$0.730 $&$\phantom{-}0.305$&$1.03$\\
$157\pm 15$&&$ 2.04\pm 0.02\pm 0.22 $&$-0.120\pm 0.032\pm 0.066 $&$0.349 $&$-0.130$&$1.02$\\
\end{tabular}
\end{ruledtabular}
\end{table*}

The measured values of the spin-averaged cross section and the
analyzing power are shown in Table~\ref{tab:ayang_table1} as
functions of the pion polar angle $\theta_q$ for three ranges in
the relative momentum $p$. The mean values and uncertainties in
the determination of $\theta_q$ have been estimated from the Monte
Carlo simulations, where possible fluctuations in the beam energy
and geometrical uncertainties in the detector system were taken
into account.

The combinations of up-down beam polarizations and left-right
detectors provide two measurements of $A_y$ as well as of the
spin-averaged cross section and these should be consistent.
However, due to the uncertainty of the energy calibration of the
$E$-counters on the two sides, the data show systematic
differences between these combinations. The deviations from the
mean value were included in the systematic error for an individual
angular bin. Compared with this, the uncertainty due to the
treatment of the non-full events as discussed in
Sec.~\ref{sec:IDPI0} is negligible for $A_y$, though it is the
dominant systematic error for the spin-averaged cross section.

The size of the $\theta_p$ angular bin is quite large
($90^{\circ}\pm30^{\circ}$) and so the analysis was repeated using
the smaller angular interval ($90^{\circ}\pm10^{\circ}$) in order
to estimate the resulting uncertainty. The results for the
spin-averaged cross section agree to within 5\%, whereas the
analyzing power results are in agreement to within the statistical
errors. The overall systematic uncertainty for the spin-averaged
cross section was estimated to be less than 16\% when the
uncertainties from the efficiency of full events and the angular
bin were taken into account. As mentioned in Sec.~\ref{sec:BPol},
the overall systematic uncertainty in $A_y$ coming from the
determination of the beam polarization is believed to be less than
6\%.

It is important to note that, due to the finite size of our
counters, the data were not taken strictly at azimuthal angles
$\varphi_q=0^{\circ}(180^{\circ})$ and the effects of this angular
spread increase when $\theta_q$ approaches
$0^{\circ}(180^{\circ})$ and $p$ gets close to its maximum allowed
value. In Table~\ref{tab:ayang_table1} we show estimates of the
average values of $\langle\cos\varphi_q\rangle$ and
$\langle\cos2\varphi_q\rangle$ evaluated using our simulation. It
is seen from Eq.~(\ref{eq:ay_rcnp}) that the product $\sigma_0A_y$
is proportional to $\cos\varphi_q$ but is independent of
$\cos2\varphi_q$ and so the estimates of
$\langle\cos\varphi_q\rangle$ have been used in
Eq.~(\ref{eq:ayex}) to deduce $A_y$. There is an explicit $\cos
2\varphi_q$ in Eq.~(\ref{eq:sigma_rcnp}), which means that the
value of $\sigma_0$ that we have measured is different from that
at $\varphi_q=0^{\circ}$. As a consequence, when dividing
$\sigma_0A_y$ by $\sigma_0$ to derive the analyzing power $A_y$,
some $\varphi_q$ dependence still remains.

To quantify the changes caused by the $\varphi_q$ variation, we
define a correction factor $d_t$ using Eq.~(\ref{eq:sigma_rcnp}),
\begin{equation}
\label{dt} d_t\equiv
\sigma_0\left(\langle\cos2\varphi_q\rangle\right)/\sigma_0(\cos2\varphi_q=1)\,.
\end{equation}
The $d_t$ given in our Table~\ref{tab:ayang_table1} were obtained
for our kinematic conditions by interpolating the coefficients
quoted in Table~IV of the IUCF work~\cite{pi0_iucf_new}. It must
be noted that the value of $H^{00}_{3}$ appearing in
Eq.~(\ref{eq:sigma_rcnp}) could not be determined in the IUCF work
and the value is assumed to be zero in the analysis. The
calculations show that the cross section increases as $\varphi_q$
moves away from $0^{\circ}$ and hence that the analyzing power
decreases.

The spin-averaged cross sections and analyzing powers found by
integrating over the polar angles $\theta_q$ are shown as
functions of the relative momentum $p$ in
Table~\ref{tab:ayint_table}. The mean values and uncertainties in
the determination of $p$ have been estimated in the same way as
for the pion polar angle $\theta_q$ in
Table~\ref{tab:ayang_table1}. The  spin-averaged cross sections
and analyzing powers have been obtained at average values of
$\cos2\varphi_q$ which depend upon $\cos\theta_q$ and they must be
subjected to a correction factor $d_t$, similar to that given in
Table~\ref{tab:ayang_table1}, in order to extrapolate the results
to $\cos2\varphi_q=1$.

\begin{table}
\caption{\label{tab:ayint_table} The $pp{\to}pp{\pi}^0$ spin-averaged
cross section, $d\sigma/d\varphi_q d\Omega_p dp$, after
integration over $\theta_q$, presented in different bins of the
relative momentum $p$. The corresponding integrated values of the
analyzing power $A_y$ are also given. The first and second error
bars denote the statistical and systematic uncertainties,
respectively, whereas that on the relative momentum $p$ is the
standard deviation that includes the effects of the binning. The
observables were measured over a range of $\varphi_q$ and, in
order to get estimates of the values at $\cos2\varphi_q=1$, the
cross section should be divided by the correction factor $d_t$ and
the analyzing power multiplied by the same factor, as discussed
for Table~\ref{tab:ayang_table1}.\\
}
\begin{ruledtabular}
\begin{tabular}{cccc}
$p$& $d\sigma/d\varphi_q d\Omega_p dp$& $A_y$ & $d_t$\\
(MeV/$c$)& \big(nb\,(rad\,sr\,MeV/$c$)$^{-1}$\big)&&\\ \hline
$ 155\pm 4$&$12.33\pm 0.04\pm 1.37 $&$-0.278\pm 0.005\pm 0.014 $&$ 1.02$\\
$ 165\pm 4$&$12.39\pm 0.03\pm 0.28 $&$-0.283\pm 0.004\pm 0.018 $&$ 1.02$\\
$ 175\pm 4$&$10.88\pm 0.03\pm 0.24 $&$-0.266\pm 0.004\pm 0.011 $&$ 1.02$\\
$ 185\pm 4$&$\phantom{1}9.52\pm 0.02\pm 0.13 $&$-0.251\pm 0.004\pm 0.005 $&$ 1.02$\\
$ 195\pm 4$&$ \phantom{1}7.42\pm 0.02\pm 0.19 $&$-0.223\pm 0.004\pm 0.010 $&$ 1.03$\\
$ 205\pm 4$&$ \phantom{1}4.76\pm 0.02\pm 0.15 $&$-0.203\pm 0.006\pm 0.011 $&$ 1.04$\\
$ 214\pm 4$&$ \phantom{1}2.12\pm 0.01\pm 0.13 $&$-0.136\pm 0.010\pm 0.040 $&$ 1.03$\\
\end{tabular}
\end{ruledtabular}
\end{table}
%
%
\section{Discussion}
\label{discussion}
%
%
\subsection{Angular dependence}
\begin{figure}[hbt]
\includegraphics[clip,width=0.45\textwidth]{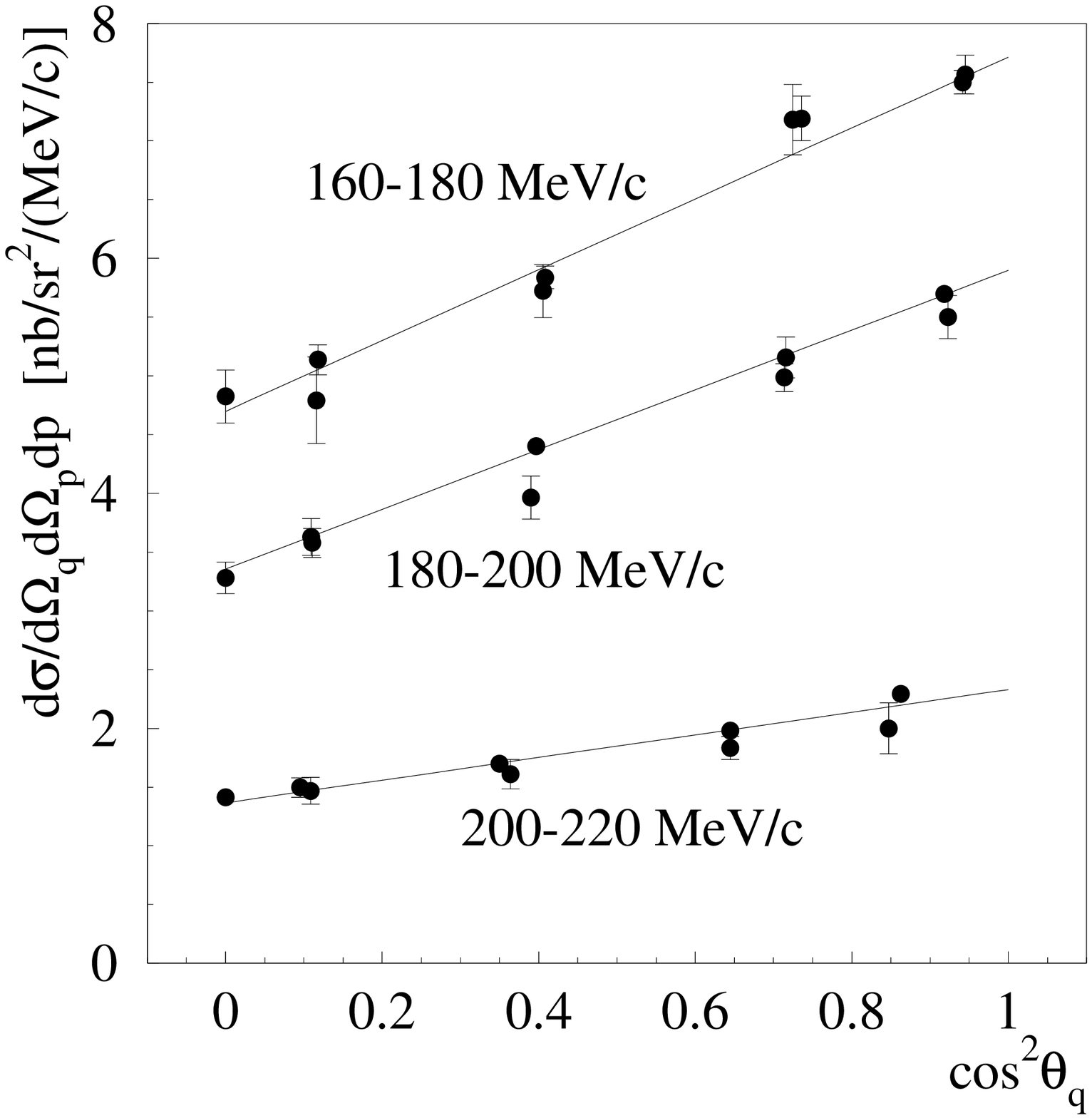}
\caption{\label{fig:crang} The angular dependence of the
spin-averaged $pp{\to}pp{\pi}^0$ cross section for the three regions
of relative momentum indicated. The error bars include the
statistical and systematic uncertainties. The solid lines show the
results of fitting the data with the linear form of
Eq.~(\ref{sig_fit}), with the resulting parameters being given in
Table~\ref{tab:angfitpara}.}
\end{figure}

The variation of the spin-averaged cross section with pion angle
is shown in Fig.~\ref{fig:crang} for the three momentum ranges.
Since, for this purpose, the two initial protons are identical,
the data are presented as functions of $\cos^2\theta_q$. The
variation seems to be linear, as expected on the basis of
Eq.~(\ref{eq:sigma_rcnp}) and the data were therefore fit with the
form
\begin{equation}
\label{sig_fit} \sigma_0=\alpha\left(1+c\cos^2\theta_q\right)\,.
\end{equation}
The resulting parameters with and without the $d_t$ factor are
given in Table~\ref{tab:angfitpara}. In all cases the
$\chi^2$/d.o.f.\ were close to unity. The results, with and
without the $d_t$ modification, are both consistent with $c$ being
constant.

The angular distribution of the pion has been investigated by
several experimental groups in the 400\,MeV region, though with
somewhat conflicting conclusions regarding the slope parameter
$c$. Thus $c=0.28\pm0.20$ was found at IUCF~\cite{pi0_iucf_new},
$0.19\pm0.01$ at TSL (PROMICE-WASA)~\cite{pi0_tsl_ang},
$-0.35\pm0.03$ at TSL (WASA)~\cite{Pia}, and $-0.19\pm0.02$ at
COSY-TOF~\cite{TOF2}. In the last of these experiments, a
dependence of the parameter upon the relative momentum could be
established, with $c$ changing from negative to positive as $p$
increases. However, even the TOF result of $0.09\pm0.01$ obtained
for $p>160\,$MeV/$c$ is significantly smaller than ours for a
similar momentum range. This difference can be explained in terms
of the different geometries of the two experiments.

\begin{table*}
\caption{\label{tab:angfitpara} Parameters extracted by fitting
the angular dependence of $\sigma_0(\theta_q)$ and
$\sigma_0A_y(\theta_q)$ with Eqs.~(\ref{sig_fit}) and
(\ref{eq:aypartial}), respectively. The values in brackets show
the results obtained when the correction factor $d_t$ is included
in the fitting process. }
\begin{ruledtabular}
\begin{tabular}{ccccc}
$p$& $\alpha$ & $c$ & $a$ & $b/a$\\
(MeV/$c$)&\big(nb/({sr}$^2$\,MeV/$c$)\big)&&\big(nb/(sr$^2$\,MeV/$c$)\big)&\\
\hline
 $160-180$&$4.82\pm0.10$ ($4.70\pm0.10) $&$0.62\pm0.04$ ($0.64\pm0.05$) &$-2.00\pm0.12$&$0.05\pm0.06$\\
 $180-200$&$3.47\pm0.05$ ($3.35\pm0.06) $&$0.72\pm0.03$ ($0.76\pm0.04$) &$-1.23\pm0.06$&$0.04\pm0.07$\\
 $200-220$&$1.44\pm0.04$ ($1.37\pm0.04) $&$0.64\pm0.08$ ($0.71\pm0.09$) &$-0.40\pm0.03$&$0.06\pm0.11$\\
\end{tabular}
\end{ruledtabular}
\end{table*}

\begin{figure}
\includegraphics[clip,width=0.3\textwidth]{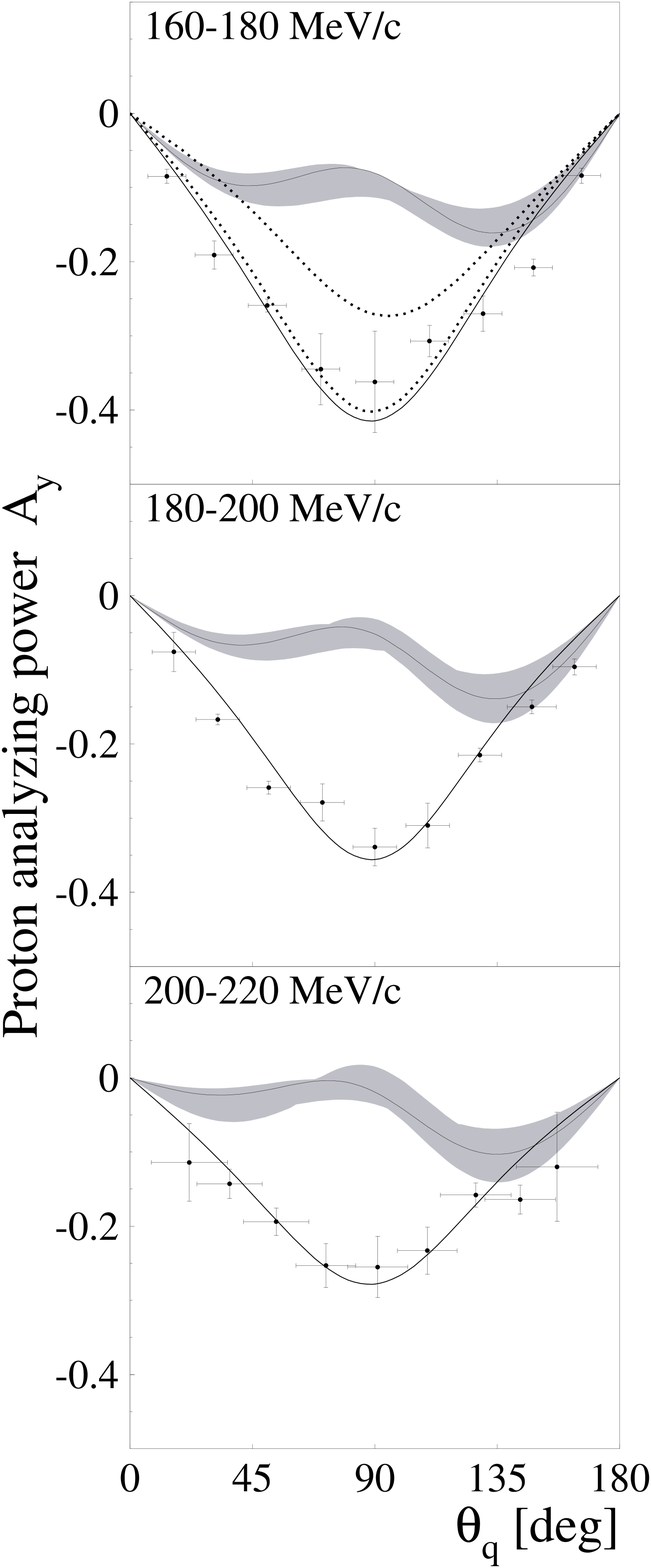}
\caption{\label{fig:ayang} The dependence of the
$\roarrow{p}p{\to}pp{\pi}^0$ analyzing power on the pion polar
production angle $\theta_q$ for the three regions of relative
momenta of the final protons. The error bars include the
statistical and systematic uncertainties. The solid curves are
fits of Eqs.~(\ref{sig_fit}) and (\ref{eq:aypartial}) to these and
the data of Fig.~\ref{fig:crang}, with the resulting parameters
being given in Table~\ref{tab:angfitpara}. The upper and lower
dotted curves in the lowest momentum region have been calculated
from Eqs.~(\ref{eq:sigma_rcnp}) and (\ref{eq:ay_rcnp}) using the
coefficients given in Ref.~\cite{pi0_iucf_new} at 375 and
400\,MeV, respectively. The curves bordered by shaded areas are
the theoretical results from the Osaka group, with the shading
indicating the expected uncertainties in the
calculation~\cite{tamura}. }
\end{figure}

Whereas at COSY-TOF the full phase space was explored, our data
were taken in essentially coplanar conditions, \emph{i.e.}\
$\varphi_p = \varphi_q = 0^{\circ}$, and with the proton polar
angle of $\theta_p = 90^{\circ}$. Using Eq.~(\ref{eq:sigma_rcnp})
with the coefficients taken from Table~IV of the IUCF
work~\cite{pi0_iucf_new}, one can derive a relation between the
slope parameter ($c$) for full acceptance and the value
($c^{\star}$) to be found with our limited coverage. Since
$H^{00}_5$ is negative~\cite{pi0_iucf_new}, $c^{\star}$ is always
larger than $c$ and for $p>160\,$MeV/$c$ one expects that
$c^{\star}\approx0.3+3.0c$. Using this relation with the COSY-TOF
value of $c$, a value of $c^{\star}=0.57\pm0.03$ is predicted and
this is consistent with our results.

The sign of the slope parameter $c$ is positive for high $p$,
reflecting the importance there of the $Pp$ contribution. On the
other hand, both the TSL~\cite{pi0_tsl_ang} and
COSY-TOF~\cite{TOF2} data show that $c$ is negative for
$p<53\,$MeV/$c$, probably due to a $SsSd$ interference that falls
very fast with increasing $p$.

Figure~\ref{fig:ayang} presents the dependence of $A_y$ on the
pion polar angle for the three relative-momentum regions. The
solid line shows the fits resulting from using the general angular
dependence of Eq.~(\ref{eq:ay_rcnp}), \textit{viz.}
\begin{equation}
\sigma_0\,A_y = a\sin\theta_q+b\sin2\theta_q\,,
\label{eq:aypartial}
\end{equation}
where the values of the free parameters $a$ and $b$ are listed in
Table~\ref{tab:angfitpara}. In the fitting, the uncertainty in the
angular determination has been included in the $\chi^2$
minimization.

It is seen from Table~\ref{tab:angfitpara} that $b$ is consistent
with zero for all three momentum bins. Since the parameters $a$
and $b$ correspond, respectively, to the $PsPp$ and ($SsSd$,
$PpPp$) contributions in Eq.~(\ref{eq:ay_rcnp}), this shows
clearly that, for the large values of $p$ investigated in our
experiment, the $PsPp$ interference terms dominate.

If the values of all the coefficients given in Table~IV of
Ref.~\cite{pi0_iucf_new} are inserted into
Eqs.~(\ref{eq:sigma_rcnp}) and (\ref{eq:ay_rcnp}), these predict
the angular dependence shown in Fig.~\ref{fig:ayang}a. The upper
and lower dotted lines are obtained by using the numbers
corresponding to 375 and 400\,MeV, respectively. It must be noted
that the IUCF parameters were obtained from averages of $A_y$ over
the full range of allowed proton-proton relative
momenta~\cite{pi0_iucf_new}. The close agreement found in the
160-180\,MeV/$c$ bin when using the 400\,MeV coefficients is due,
in part, to the cross section being maximal in this region.

In contrast, the model calculations of the Osaka
group~\cite{tamura}, shown in Fig.~\ref{fig:ayang}, demonstrate a
much more asymmetrical behavior than the experimental data and
greatly underestimate the magnitude of the analyzing power. The
predictions for the $Pp$ or $Ps$ transitions are therefore much
smaller than those found experimentally. More theoretical studies
are needed in order to elucidate the origin of this disagreement.
%
%
%
\subsection{Momentum dependence}\label{Momentumdependence}
\begin{figure}
\includegraphics[clip,width=0.40\textwidth]{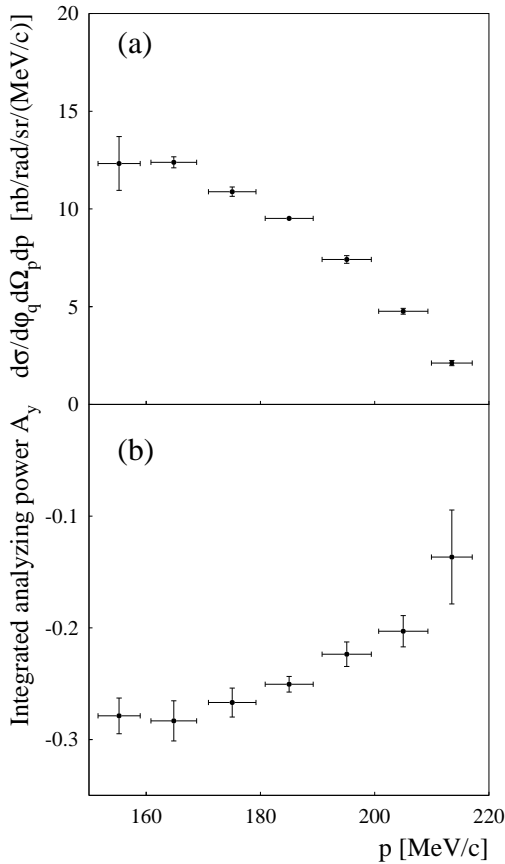}
\caption{\label{fig:ayint} (a) Spin-averaged cross section as a
function of the relative momentum of the final protons. (b) The
relative momentum dependence of the integrated analyzing power.
The error bars include the statistical and systematic
uncertainties. }
\end{figure}

Figures~\ref{fig:ayint}a and \ref{fig:ayint}b show, respectively,
the spin-averaged cross section and analyzing power integrated
over $\theta_q$ as functions of the relative momentum of the final
protons. The error bars include both systematic and statistical
uncertainties. On kinematic grounds, $A_y$ must tend to zero as
$p$ approaches the highest allowed values because the pion
momentum vanishes in this limit. As previously remarked, the cross
section seems to be maximal at our lowest values of $p$, as does
the magnitude of the analyzing power.
%
%
\section{Phenomenological analysis}
\label{phenomenal}
It was already stressed that the combination $\sigma_0A_y$ is not
sensitive to the $\cos2\varphi_q$ term in
Eq.~(\ref{eq:sigma_rcnp}) and so does not suffer from the
resultant $d_t$ ambiguity. Since the integral of the
$\sin2\theta_q$ term in Eq.~(\ref{eq:aypartial}) over the pion
angle vanishes, the dependence of the integrated $\sigma_0A_y(p)$
can be expressed in terms of the interference between the two
types of $P$-state amplitudes multiplied by the three-body
phase-space factor $\rho(p)$;
\begin{equation}
\sigma_0A_y(p) = PsPp \:\rho(p). \label{eq:aymon}
\end{equation}

The $PsPp$ corresponds to a linear combination of the coefficients
$G^{y0}_{1,2,4}$ in Eq.~(\ref{eq:ay_rcnp}). According to
Ref.~\cite{pi0_iucf_new}, $G^{y0}_1$, which does not contain any
$\theta_p$ dependence, contributes about 95\% of the total
magnitude of $A_y$ so that to a good approximation the
$G^{y0}_{2,4}$ may be neglected. In this case, $PsPp$ results from
$s$-$p$ interferences with the same $P$-states of the final
protons. It can therefore
be expressed as
\begin{equation}
PsPp = C_0\,f_{{^{3\!}{P}_0}s}\:f_{{^{3\!}{P}_0}p} +
C_2\,f_{{^{3\!}{P}_2}s}\:f_{{^{3\!}{P}_2}p}, \label{eq:PsPp}
\end{equation}
where $C_{0,2}$ are parameters and $f_{{^{3\!}{P}_{0,2}}s}$ and
$f_{{^{3\!}{P}_{0,2}}p}$ represent the pion $s$- and $p$-wave
amplitudes, respectively.

A partial-wave amplitude for pion production may be approximated
in terms of the overlap integral involving the pion-production
operators $\hat{\Pi}(r)$ and the radial wave function of initial
$u_i(r)$, and final $u_f^{p}(r)$ protons and the pion plane wave
$j_l(\fmn{1}{2}qr)$,
\begin{equation}
f_{f l}=\int u_f^{p}(r)\,j_l(\fmn{1}{2}qr)\,\hat{\Pi}(r)\,
u_i(r)\,r^2\,d r, \label{eq:overlap}
\end{equation}
where $r$ is the proton-proton separation distance, and $l$ and
$q$ is the pion angular momentum and momentum, respectively.
Analytic forms for the $Ss$-wave pion-production amplitude may be
found in Ref.~\cite{Horowitz}.

If the pion-production operator is taken to be
momentum-independent, any strong momentum dependence of the
partial-wave amplitude must be ascribed to the variation of the
radial wave function of the final protons and pion with $p$ and
$q$. To investigate this further, we approximate
Eq.~(\ref{eq:overlap}) by evaluating the product of the wave
functions at a \textit{fixed}
distance $r=R$, in which case %
\begin{equation}
f_{f l} \propto{}u_f^{p}(R)\,j_l(\fmn{1}{2}qR).
\label{eq:appromatrix}
\end{equation}

\begin{figure}
\includegraphics[clip,width=0.40\textwidth]{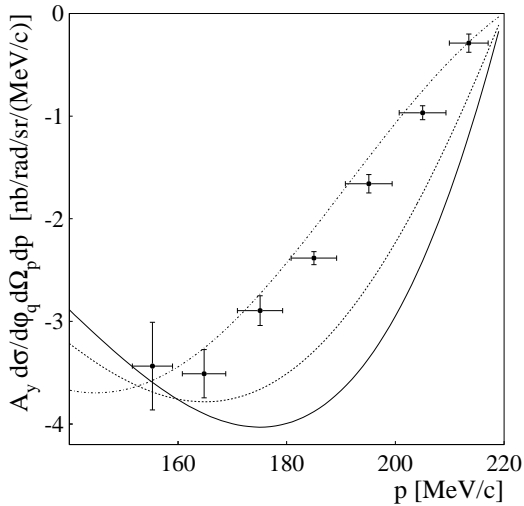}
\caption{\label{fig:PsPp} The relative momentum dependence of the
angle-integrated $A_y\sigma_0(p)$. The error bars include the
statistical and systematic uncertainties. The lines show the
variation predicted for the $^{3\!}P_2$ state when using
Eq.~(\ref{eq:appromatrix}) with different values of the
interaction range. The solid line shows the calculation with
$R=1\,$fm, whereas the dashed line and dot-dashed line show the
calculations with $R=2\,$fm and $R=3\,$fm, respectively. }
\end{figure}

The $^{3\!}P_{0,2}$ radial wave functions have been derived from
the Paris potential with the Coulomb interaction~\cite{Paris}.
Since the experimental data cannot distinguish between the
contributions from the ${^{3\!}P_{0}}$ and ${^{3\!}P_{2}}$ states,
the calculations have been carried out taking them into account
one at a time. However, there is very little difference between
the two sets of results and only those for the $^{3\!}P_{2}$
states are shown in Fig.~\ref{fig:PsPp}. The overall normalization
factor $C_{0,2}$ has been fixed on the basis of
Eq.~(\ref{eq:ay_rcnp}) to satisfy $\int \sigma_0 A_y(p) dp =
\sigma_{\text{tot}}\,(G^{y0}_1-G^{y0}_2+G^{y0}_4)/16{\pi}$, where
the value $\sigma_{\text{tot}}=65\,\mu$b has been obtained by
interpolation of the IUCF data~\cite{pi0_iucf_new}.

The $\sigma_0 A_y$ data are only sensitive to a value of the
distance  $R=R_{Ps}=R_{Pp}$. As demonstrated in
Fig.~\ref{fig:PsPp}, the curves obtained from
Eq.~(\ref{eq:appromatrix}) using radii $R=1$, $R=2$, and $R=3\,$fm
show significantly different dependences on the relative momentum.
The $R=1$\,fm line fails to reproduce the data but, as $R$ is
increased, the minimum in the curve moves towards lower momenta
and the data are better described. Within the framework of this
simplistic analysis, the data seem to require a pion-production
operator with a fairly long range component. It was noted for the
TSL measurements of the differential cross section
$d\sigma/dp$~\cite{pi0_tsl_ang} that the momentum dependence could
be described by taking the exchanged particle in
Eq.~(\ref{eq:overlap}) to have the mass of the pion but that such
a description was no longer possible when the $\rho$-meson mass
was used.
%
%
\section{Summary}
\label{summary}
We have measured the angular and momentum dependence of the cross
section and analyzing power of the $\roarrow{p}p{\to}pp\pi^0$
reaction at an incident energy of 390\,MeV. Recoil protons stopped
in a scintillation counter have been identified by the
$\Delta{E}$-$E$ technique. Though the counters in the horizontal
plane covered angles between $15^{\circ}$ to $35^{\circ}$ to the
left and right of the beam, the vertical acceptance was quite
small and this feature is crucial in any theoretical description
of the data. Only high relative momentum $p>160\,$MeV/$c$ were
registered and, under these conditions, final $P$-state proton
pairs play an important role.

Although the dependence of the cross section on the pion angle
shows a larger anisotropy than in the case of data obtained with
almost full acceptance~\cite{pi0_iucf_new,pi0_tsl_ang,TOF2}, this
is in large part the result of our coplanar geometry. After taking
this into account by using the IUCF
parameters~\cite{pi0_iucf_new}, our results do not contradict
those recently published by COSY-TOF~\cite{TOF2}.

The shape of $\sigma_0A_y(\theta_q)$ follows quite closely a
$\sin\theta_q$ form, which is consistent with this observable
being dominated by $PpPs$ interference. The contribution from the
$SsSd$ term is negligible at the high relative momentum studied in
this experiment. However, the behavior of $A_y$ at small $p$ could
be a useful probe to investigate the role of the $SsSd$
contribution~\cite{ANKEPRO}.

The momentum dependence of the analyzing power at large $p$ is
also consistent with the $PpPs$ interference interpretation. The
variation can be explained within a very simple model by taking
the pion-production operator to have a rather long range.

Our results, taken in conjunction with the IUCF double-polarized
data~\cite{pi0_iucf_new} and other published results on the
unpolarized cross sections, may provide the extra information
necessary for the understanding of the production mechanisms for
$P$-state protons. However, to succeed in this, a much more
sophisticated theoretical model is required to replace the rather
qualitative approach used here to describe the data.
%
%
\section*{Acknowledgements}
The authors are grateful to the cyclotron staff for their support
throughout this experiment. We acknowledge the help of K.~Sagara
with the liquid hydrogen target system. Comments from Professor
A.~Johansson have been particularly valuable. This work was
performed at RCNP under the program E140.
%
%

%
\end{document}